\documentclass[journal,10pt]{IEEEtran}
\IEEEoverridecommandlockouts
\usepackage{cite}
\usepackage[utf8]{inputenc}
\usepackage{enumerate}
\usepackage[caption=false,font=footnotesize]{subfig}
\usepackage[ruled,linesnumbered]{algorithm2e}
\SetKwComment{Comment}{/* }{ */}
\SetKwInput{kwInput}{Given}

\usepackage{amsmath}
\usepackage{amsfonts}
\usepackage{amsthm}
\usepackage{threeparttable}
\usepackage{pdfpages}
\usepackage{multirow}
\usepackage{hyperref}
\DeclareMathOperator*{\argmin}{arg\,min}

\title{Designing Discontinuities\thanks{The authors are with the Coordinated Science Laboratory, University of Illinois Urbana-Champaign, Urbana, IL  61801, USA (e-mail: \{iferwna2, varshney\}@illinois.edu). 
 IF is also with the Department of Computer Science and LRV is also with the Department of Electrical and Computer Engineering.}\thanks{This work was supported in part by NSF grant CCF-1717530, by the SURGE Graduate Fellowship, and by the Center for Pathogen Diagnostics through the ZJU-UIUC Dynamic Engineering Science Interdisciplinary Research Enterprise (DESIRE).}} 
\author{Ibtihal Ferwana, Suyoung Park, Ting-Yi Wu, and
Lav R.\ Varshney, \IEEEmembership{IEEE Senior Member}}

\begin{document}

\maketitle
\begin{abstract}
Discontinuities can be fairly arbitrary but also cause a significant impact on outcomes in larger systems.  Indeed, their arbitrariness is why they have been used to infer causal relationships among variables in numerous settings. Regression discontinuity from econometrics assumes the existence of a discontinuous variable that splits the population into distinct partitions to estimate the causal effects of a given phenomenon. Here we consider the \emph{design} of partitions for a given discontinuous variable to optimize a certain effect previously studied using regression discontinuity. To do so, we propose a quantization-theoretic approach to optimize the effect of interest, first learning the causal effect size of a given discontinuous variable and then applying dynamic programming for optimal quantization design of discontinuities to balance the gain and loss in that effect size. We also develop a computationally-efficient reinforcement learning algorithm for the dynamic programming formulation of optimal quantization. We demonstrate our approach by designing optimal time zone borders for counterfactuals of social capital, social mobility, and health. This is based on regression discontinuity analyses we perform on novel data, which may be of independent empirical interest.   
\end{abstract}
 
\section{Introduction}
Quantization is a central problem in signal processing and causal inference is an emerging one.  Their intersection  has remained unstudied, but here we consider quantizer design to optimize inferred causal effects.

Whether one earns admission to a particular school on the basis of a test score is often linked to significant educational and life outcomes \cite{ParkSHA2015, EstradaG2017}, but the admissions threshold may be quite arbitrary. Those on one side of the threshold may otherwise be quite similar to those on the other side. Whether one receives an important health intervention on the basis of a fairly arbitrary blood pressure threshold is linked to significant differences in health outcomes and mortality; again patients on either side of the threshold may be largely similar otherwise \cite{VenkataramaniBJ2016}. There are similarly arbitrary discontinuities in numerous settings in public policy, economics, healthcare, engineering systems, and elsewhere that may cause significant impacts. Indeed, categorization on the basis of fairly arbitrary partitioning of certain attributes abounds in life \cite{VarshneyV2017}.

Discontinuity---the presence of a discrete set of partitions---has been used to learn causal relationships among variables. Indeed a leading method for causal inference in econometrics is regression discontinuity design (RDD), a quasi-experimental method that assumes a pre-known threshold dividing the population into two discontinuous groups \cite{AngristP2009}. Comparing samples on each side of the threshold then allows inference of causality. The basic assumption is that all potentially relevant variables besides the treatment variable and outcome variable are continuous at the point where the treatment and outcome discontinuities occur \cite{CattaneoIT2020}.

RDD has been applied in various domains with discontinuities. For example, car age and mileage have both been used as discontinuous measures represented by a set of thresholds for used car prices \cite{englmaier2018price}. In national security, the causal effects of air strikes on violence control were estimated when the variable of air strikes was  discontinuous as a function of geographical location and local economic characteristics \cite{dell2018nation}.

These examples apply RDD by comparing samples on either side of a single or several predefined thresholds. However, there are settings where the choice of cutoff point can change the estimated effect for the same variable. That is, some settings might allow the \emph{(re)design} of thresholds to create different discontinuities, and hence be subject to the effects of different partitions. For example in faculty performance reviews, gender bias may be reduced by moving from a 10-point evaluation scale to a 6-point scale \cite{rivera2019reviews}. 

Here, we take up the challenge of partition design in the context of causal effects of discontinuities, which as far as we know has not been studied in any systematic manner.  Indeed, there appears to be a lacuna at this intersection of causal inference and mechanism design approaches from signal processing (or any other approach, for that matter).  To do so, we first review and slightly extend ways to address the counterfactuals of redesigning discontinuities.  Then we develop new techniques in optimal quantization theory for partition design.  Although there is a large literature in quantization theory \cite{GrayN1998, GershoG1991}---including in settings of statistical inference \cite{PoorT1977, VarshneyV2008, MisraGV2011, ShlezingerER2019}---there appears to be no prior work on quantization in the context of causal inference.

A common approach to optimal quantization is the Lloyd-Max algorithm \cite{max1960quantizing, lloyd1960least}, which finds optimal partitions based on centroids, so a point belongs to a partition based on its distance from the center. However, we aim to design partitions based on minimizing the distance between points (within a partition) and the boundaries of the partition. Specifically, as we will see, we aim to concentrate the probability mass at the boundaries of partition regions, rather than largely near the center. Therefore, the Lloyd-Max algorithm does not work in our setting. Instead we extend a dynamic programming algorithm for optimal quantization  \cite{sharma1978design}. We further develop a computationally-efficient reinforcement learning algorithm for partition design.

\subsection{Time Zone Discontinuity}
Time zones are an example of discontinuities that affect daily life. In the late 19th century, the 24 time zone system emerged so nearby places would share a common timekeeping standard. The standard time zone system was adopted globally, perhaps most notably by various railroad companies across the United States \cite{zerubavel1982standardization}. Normatively, the 720 degrees of longitude on Earth are divided into 24 time zones: traveling from east to west, one hour is lost every 15 degrees. However, there are countries that geographically lie in certain time zones but actually follow other time zones, such as France and Spain, which use Western European Time. China geographically spans five time zones but uses just a single China Standard Time throughout. Time zone borders may not run straight from north to south, but may follow certain political boundaries. 

Although it may seem fanciful to consider the redesign of time zones given the  international regulatory and standards regime that would need to be changed, we take it as our running example throughout the paper.  As we will see, the causal impacts of time zone boundaries on human wellbeing can be large, when compared to major technology diffusions \cite{NunnQ2011}.  One should also note that the Prime Meridian in Greenwich, England that defines the modern time zone system is quite arbitrary \cite{Dohrn1998} and there have been historical alternatives such as Paris and Ujjain \cite{Burgess1860}. 

Why should time zone boundaries have an impact on societal outcomes?  
Time zone systems along with globalized social systems led to uniform  start times of school, work, and sleep, which may dictate wake-up times not aligned with the sun \cite{hamermesh2008cues}. Such schedules can disrupt human circadian rhythms and have consequences on health and productivity \cite{cappuccio2010sleep}. 

The timing of natural light at a given location is determined by time zone borders, which has direct effects on the sleep-wake cycle. At locations on the eastern (right) side of a time zone border, the sun sets an hour later than in nearby locations on the opposite side of the border. By exploiting the discontinuity in the timing of natural light at time zone boundaries, \cite{Giuntella_2019} found that an extra hour of natural light in the evening reduces sleep duration by approximately 20 minutes. Similarly, \cite{jagnani2018poor} found  later sunset reduces a child's sleep time, which in turn affected educational outcomes \cite{jagnani2018poor}.  Likewise, there have been findings of impacts on productivity and earnings: notably, causal inference shows a one-hour increase in location-average weekly sleep increases earnings by $1.1$\% in the short run and $5$\% in the long run \cite{gibson2018time}.  There are numerous other examples of regression discontinuity designs on the basis of time zone boundaries, e.g.\ \cite{SchaferH2020, HeisselN2018, HolbeinSD2019, Freed2023, Ali2023}. 

Not only do time zone boundaries influence sleep, but also increased exposure to sunlight increases the misalignment between social time and internal circadian timing. Disruption in circadian rhythm was found to increase various health risks for specific cancers \cite{gu2017longitude}. These results imply sunset time, as linked to geographic location, may contribute in other ways to health and wellbeing. 

Here, we ask how to partition the world into time zones that optimize human health, social capital, and social mobility, rather than the arbitrary Greenwich system in place today. To re-emphasize, we need two things to do this: first, the counterfactual prediction using regression discontinuity to measure the effect of current time zone borders, and second, quanitization to design the optimal time zone borders (optimal time zone discontinuities). 

\subsection{Organization}
The remainder of this paper is organized as follows.  Section~\ref{sec:background} provides background on causal inference, including regression discontinuity, and a small extension on counterfactual prediction. Section~\ref{sec:quatizer} provides the details for the quantization formulation we develop, including computational techniques for dynamic programming and reinforcement learning for solving the quantization objective. Section~\ref{sec:empirical} has new empirical investigations we conduct, first  introducing datasets and pre-processing, then reporting results of counterfactual prediction, the outputs of the redesigned time zone borders, and finally the causal inference results under redesigned time zone borders. The empirical results on social capital may be of independent interest. 
 Lastly, Section~\ref{sec:discussion} concludes with a discussion.  

\section{Causal Inference}
\label{sec:background}
In engineering, health, and social science fields, the randomized experiment has played important roles to uncover causal effects under a given intervention on an outcome of interest. For example, to study the efficacy of a new drug, one can randomly assign patients to two groups where one group receives the new drug and the other  receives a placebo. By comparing the difference in efficacy between the two groups, one can estimate the treatment effect. Although randomized experiments have been robust in producing estimates and simple interpretations, they are often hard to apply to real-world applications given practical and ethical limitations in randomly assigning groups and interventions. 
To overcome such limitations, non-experimental designs have been developed to uncover causal effects.

One of the main distinctions between standard statistical analysis and causal inference is dealing with changing conditions. Statistical analysis, represented by regression and hypothesis testing techniques,  estimates beliefs from past to future as long as experiment conditions do not change, whereas causal inference infers beliefs under changing conditions to uncover causal relationships among variables \cite{pearl2009causalInferenc}. Several frameworks are used for causality analysis, such as structural models \cite{pearl2010causal} and the potential outcome framework \cite{rubin1974causaleffect}, which we focus on here. 

The potential outcome framework assumes effects are tied to a treatment or an intervention. To reveal the causal effects of an intervention, \cite{rubin1974causaleffect} proposed to measure the difference between two potential outcomes; let us denote them as $Y^N$ and $Y^I$, for a given unit $x$. The potential outcome $Y^N$ is the outcome for $x$ without being exposed to an intervention, and $Y^I$ is the outcome after an intervention is applied on $x$. So, the causal effect is 
\begin{equation}
    \tau = Y^N-Y^I \mbox{ .} 
    \label{eq:effect}
\end{equation}
However, we can never observe both outcomes for the same unit under the same conditions, only one of the two will happen at a given time. Since one of the potential outcomes will always be unavailable, the core objective is to estimate it. 

Let us introduce the main terms in the potential outcomes literature, which are used throughout. A \textit{unit} is the atomic object in the framework, which can be a city or a county. A \textit{treatment} is the action applied to a unit to change its state. The treatment \footnote{The terms \emph{treatment} and \emph{intervention} are used interchangeably.} can be a medicine given to a particular group. The treatment is usually thought of as binary, so one group receives the treatment (the \textit{treated} group), and the other does not (the \textit{control} group). One  common design under the potential outcomes framework is regression discontinuity, which we consider here. 

\subsection{Regression Discontinuity}
RDD has recently gained attention because it provides a credible analysis of causal effects with relatively mild assumptions compared to other non-experimental designs such as instrumental variables. In RDD, each observation can be split into two groups based on a known discontinuous variable---a cut-off point for an intervention of interest \cite{imbens2008regression}. Suppose observations cannot perfectly manipulate the intervention; then the difference between two groups near the known cut-off point can be used to measure the local average treatment effect (LATE), which is the effect on an outcome of interest given a treatment. Therefore, RDD does not require explicit randomization of the treatment, yet it gives the comparative analysis of the causal effects as the randomized experiment. 

The causal estimate arises from the comparison between both groups, in which the distribution of samples below and above the threshold are expected to be different if an intervention had an impact on treated samples. However, to ensure the validity of the estimate, the distribution of characteristics around the threshold should not change discontinuously \cite{lee2010regression}. 

\paragraph{Model Validity} We need two assumptions to make RDDs valid. First, the individual should not be able to manipulate the assignment variable precisely. Intuitively, if the individual had a choice to manipulate the assignment variable, the difference between just above and just below the cut-off point cannot be solely explained by the assignment variable. Further, the probability of being treated (just above the cut-off point) and in the control group  (just below the cut-off point) may not be the same. 
To check the manipulation assumption, we will perform the McCrary  test \cite{McCrary_2008}. 

Second, all covariates and unobservable variables should not have a discontinuity and they should be continuous with regard to the assignment variable. Otherwise, the average treatment effect can be confounded by them and our estimates will be biased. Also, this is important for our potential outcome in counterfactual prediction because we assume that the distribution of treatment and control group should be the same over all assignment variables (i.e., $\mathbb{E}[Y_i^{I}|X]$ and $\mathbb{E}[Y_i^{N})|X]$ are continuous). To check continuity assumptions, we will investigate the continuity of observed covariates  using the covariate index \cite{Card_2012,Kumar_2018}.

\begin{figure}[!ht]
  \centering
    \includegraphics[width=0.6\linewidth]{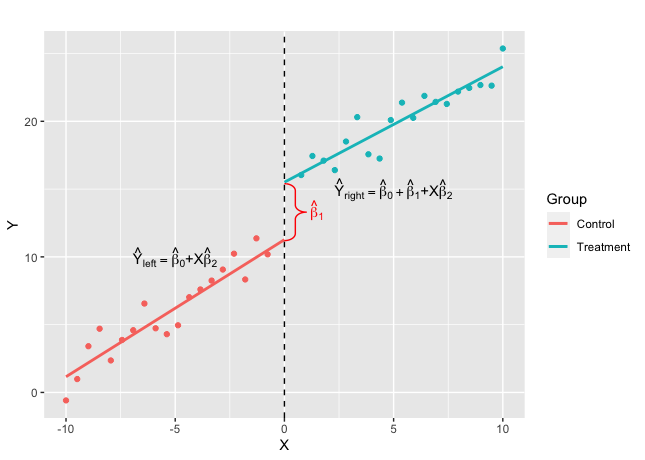}
\caption{Simple linear regression discontinuity design.}
\label{Fig:RDD_example}
\end{figure}

\paragraph{Empirical Specification} In RDD, we are interested in the average treatment effect, $\tau$. If the relationship between $X$ and $Y$ is linear, we can estimate the treatment effect $\beta_1$ by fitting the linear model:
\begin{equation}
\label{eq:Simple_RDD}
    Y = \beta_0 + h\beta_1 + X\beta_2 + \varepsilon \mbox{,}
\end{equation}
where $X$ is the assignment variable, $h \in \{0,1\}$ is the treatment variable where $h=1$ if $X \geq C$ and $h=0$ if $X < C$ where $C$ is a known cut-off point, and $\varepsilon$ is the error term in the regression model. Since we can only observe either $\mathbb{E}[Y_i|X,h=1]$ (denoted by $Y_i^I$) to the right of the cut-off point or $\mathbb{E}[Y_i|X,h=0]$ (denoted by $Y_i^N$) to the left of the cut-off point, the average treatment effect is 
\begin{equation}
\begin{split}
    & \lim_{\varepsilon	\to 0^{+}}{\mathbb{E}[Y_i|X_i=C+\varepsilon]}-\lim_{\varepsilon \to 0^{-}}{\mathbb{E}[Y_i|X_i=C+\varepsilon]}  \\ 
    & = \lim{\mathbb{E}[Y_i^I-Y_i^N|X_i=C]} = \hat{\beta_1} \mbox{.}
    \end{split}
\end{equation}
We depict this  in Figure \ref{Fig:RDD_example}. The treatment variable $D$ is determined by the assignment variable $X$ with the cut-off point at $0$. The difference between two regression lines near the cut-off point represents the LATE, $\hat{\beta}_1$.

\subsection{Counterfactual Effects}
Counterfactual prediction in this context is essentially extrapolation using the regression-based model.  How can one estimate the counterfactual of shifting discontinuities?  

Given the RDD setting under consideration, we propose to apply RDD twice on data to measure the counterfactual effect of a newly determined set of discontinuities. Specifically, we apply the RDD model on data first with the current set of discontinuities and derive causal effects. After we design optimal discontinuities using quantization theory, we again apply the RDD model with the new set of discontinuities.  Note that in both cases, the data has the same (spatial) distribution, but differs in the partitioning. Thus, we observe the change in the counterfactual effects between both cases and measure the improvement in the desired effect.

\subsection{Running Example: Causal Effects of Time Zones on Human Wellbeing}
\label{sec:RD_setting}
Consider human health, social capital, or social mobility.  To measure the discrepancy between, say social capital, when given standard time zone borders, we will apply RDD \cite{lee2010regression} to estimate causal effects. We will fit a linear model to measure the causal effects of time zone borders by modeling the distance to borders as discontinuous. In RDD, we have two groups based on a cut-off point---treated and control---and we aim to find the average treatment effect as in  \eqref{eq:effect}. Since we cannot observe two outcomes simultaneously for a given region at a given point of time, we estimate the average treatment effect to derive the causal effects on the outcome of interest. 

We will fit the following RD model:
\begin{equation}
    Y_{c} = \beta_0 + \beta_1 h_c +  \beta_2 f(d_c) + h_c * \beta_3 f(d_c) +  \beta_4 \delta_c + \varepsilon_i \mbox{,}
    \label{eq:rd}
\end{equation}
where $Y_{c}$ is the outcome variable for a region $c$, $h_{c} \in \{0,1\}$ is a dummy variable where 1 indicates the eastern side of a time zone boundary and 0 indicates the western side based on the distance $d_c$ from the time zone boundary (positive $d_c$ value is for the eastern side and negative $d_c$ value is for the western side). The function $f(\cdot)$ is taken as polynomial \cite{gelman2019high}. The control variable $\delta_c$ includes socio-demographic variables to improve the accuracy of estimates, and $\varepsilon_i$ is the error term in the regression model. The treatment effect is $\beta_1$ which measures the effect size of the distance from time zone borders on the variable of interest. We pick the coefficient $\beta_1$ to reflect the treatment effect since $\beta_1$ captures the treatment effect at the discontinuity when $d_c = 0$.

\section{Quantizer Design}
\label{sec:quatizer}
Taking the time zone design problem for concreteness, let us consider designing optimal time zone borders that maximize human wellbeing. That is, we consider the problem of partitioning the geographic distribution of the human population through the choice of time zone borders so as to optimize social capital or health. Basically, we want as many people living at the eastern edges of their time zones so the sun sets earlier, they sleep more, and therefore have better wellbeing.  So as not to have adverse circadian rhythm effects of mismatch between clock and sun, we also want to minimize the mismatch between the clock time and the solar time. In our empirical work, we will measure the gain that optimal time zone regions have on wellbeing in several ways, e.g.\ social capital, COVID-19 cases, and social mobility.

Let us consider a quantizer that partitions the set of longitude lines $\mathcal{B}$ into $K$ subsets, $\mathcal{B}_1, \dots, \mathcal{B}_K$, called quantization regions. The regions are intervals bounded by time zone borders $b^{(k)}_i$, for $K$ time zone borders, $k\in [0,K]$ and $N$ longitude lines, $i \in [0,N]$. For example, the interval for region $\mathcal{B}_1 = (-\infty,b^{(1)}_i], \mathcal{B}_2 = (b^{(1)}_i,b^{(2)}_{i^{\prime}}], \dots, \mathcal{B}_{K} = (b^{(K)}_{i^{\prime \prime}},\infty)$. For each quantization region $\mathcal{B}_k$, there is a representation point $r^{(k-1)}_i$ to which elements are mapped. The values $r^{(k-1)}_i$ are in the middle of the region $k$, and the values $b^{(k)}_i$ are at the boundaries of the region $\mathcal{B}_k$. 

To achieve our goal of having optimal partitions of the human population, we measure the effect of a given time zone border position on the rest of the longitude lines. Let $\mathcal{D}(b_j, b_i^{(k)})$ be a function that measures the distortion on the population (at longitude line $b_j$) being at the eastern side of the time zone border $b_i^{(k)}$. Similarly, let $\mathcal{D}(b_j, r_i^{(k)})$ be the distortion from the circadian mismatch, such that the population at longitude line $b_j$ are distant from the middle of the region. Therefore, we aim to minimize the distortion brought by the eastern edge effect $\mathcal{D}(b_j, b_i^{(k)})$ and the circadian rhythm effect $\mathcal{D}(b_j, r_i^{(k)})$ for the population at $b_j$ having $b_i^{(k)}$ to be their time zone border. We approach the problem in three formulations of increasing intricacy, (a)--(c). 

\paragraph{Prime Meridian Choice}
The first optimization is that we search for a longitude line $b_i^{(k^o)}$ that acts as the reference time zone border for other borders. That reference time zone border $b_i^{(k^o)}$ minimizes the average distortion $\mathcal{D}(\cdot)$ for all other longitude lines. A direct optimization is to fix $K=24$ and fix the quantizer to be uniform-sized such that regions are of size $720/24 = 30$, for $720$ longitude lines. Therefore, the objective is to minimize the distortion of the eastern edge effect 
\begin{equation}
    \argmin\limits_{b_i^{(k^{o})}} \mathbb{E}\left[\sum\limits_{j=0}^{N}\mathcal{D}(b_j, b_i^{(k)})\right] \mbox{.}
    \label{eq:prime_meridian}
\end{equation}

\paragraph{Time Zone Boundaries Choice}
In the second optimization, we allow regions to be of non-uniform size and search for the optimal boundaries of each time zone region. Given that the Earth is to be divided into 24 regions, for each time zone, we set $K=24$ and find the optimal $K$ partitions. For each $k\in K$, we find the time zone border $b_i^{(k)}$ and the corresponding representation point $r_{i^{\prime}}^{(k)}$ that minimize the distortion at longitude lines in partition $\mathcal{B}_{k}$. Therefore, the objective is to find the set of optimal $K$ time zone borders, $\{b^{(1)}_i,\dots, b^{(K)}_{i^{\prime}}\}$, and their corresponding representation (middle) points $\{r^{(0)}_n,\dots, r^{(K-1)}_{n^{\prime}}\}$. 
\begin{equation}
\begin{split}
    \argmin\limits_{\{b^{(1)}_i,\dots, b^{(K)}_{i^{\prime}}\},\{r^{(0)}_n,\dots, r^{(K-1)}_{n^{\prime}}\}} 
    \mathbb{E}\left[\sum\limits_{j=0}^{N}\mathcal{D}(r_j,r^{(k)}_n)\right] \\
    + \lambda \mathbb{E}\left[\sum\limits_{j=0}^{N}\mathcal{D}(b_j,b^{(k)}_i)\right]\mbox{,}
\end{split}
    \label{eq:timezone_choice}
\end{equation}
where $i, i^{\prime}, j, n, n^{\prime} \in [0, N]$. The quantity $\lambda$ is meant to reflect the causal impact of current timezone borders on wellbeing. Therefore, the effect estimated, $\beta_1$ in \eqref{eq:rd}, from the current time zone borders on wellbeing is substituted by $\lambda$ within the quantization formulation.

\paragraph{Numbers and Boundaries of Time Zones}
The last formulation makes $K$ open to optimization to find the optimal number of time zone borders, given the constraints on the partition boundaries and representation points. There is a cost associated with more time zones, weighted by $\eta$, since that can make global coordination more difficult (indeed the initial motivation for standard time zones was to facilitate coordination).
\begin{equation}
\begin{split}
    \argmin\limits_{K, \{b^{(1)}_i,\dots, b^{(K)}_{i^{\prime}}\},\{r^{(0)}_n,\dots, r^{(K-1)}_{n^{\prime}}\}} 
    & \mathbb{E}\left[\sum\limits_{j=0}^{N}\mathcal{D}(r_j,r^{(k)}_n)\right] \\
    & + \lambda \mathbb{E}\left[\sum\limits_{j=0}^{N}\mathcal{D}(b_j,b^{(k)}_i)\right] \\ 
    & + \eta K  \mbox{.}
\end{split}
    \label{eq:k_opt}
\end{equation} 

\subsection{Dynamic Programming Formulation}
The first optimization in \eqref{eq:prime_meridian} can be approached by exhaustive enumeration whereas finding optimal partitions using \eqref{eq:timezone_choice} or \eqref{eq:k_opt} is approached through dynamic programming (DP).\footnote{For brevity, we focus the paper on \eqref{eq:timezone_choice} rather than \eqref{eq:k_opt}.} Indeed, we extend a DP algorithm for optimal quantization  \cite{sharma1978design}. A common approach for quantization---the Lloyd-Max algorithm \cite{max1960quantizing, lloyd1960least}---does not work in our setting, since we aim to concentrate the probability mass at the right (eastern) edge of partition regions, $b_i$ rather than largely being near the center $r_i$. 

Recall $N$ is the total number of longitude lines, with $K$ time zone borders. We aim to find the optimal $K$-level quantizer, where $b_i^{(k)}$ acts as the time zone border of minimum distortion for region $\mathcal{B}_k$. 

Specifically to measure the distortion $\mathcal{D}(\cdot)$, consider two longitude lines, $b_i$ and $b_j$, where $b_i$ acts as a reference time zone border for $b_j$. We calculate the eastern edge effect to be the amount of population at longitude lines to the eastern side of $b_i$ up to $b_j$. We start by measuring the eastern edge effect $\mathcal{D}_k(b_i, b_j)$ at just one quantization level, $k=1$, i.e.\ one time zone region, assuming longitude lines belong to one segment as follows:
\begin{equation}
    \mathcal{D}_1(b_i,b_j) = 
    \sum_{w=i}^{j} \rho Z_w \mbox{,}
    \label{eq:d1}
\end{equation}
where $Z_w$ is the population size at longitude line $b_w$, and $\rho$ is a scaling factor, i.e.\ $0.5 \times (|i-w|)$, to reflect the distance between $b_i$ and $b_j$.  

Next, we define the distortion at $2\leq k \leq K$, when $k$ segments are placed in the interval $(b_i, b_j)$. For each $k$, we define a value $M_k = N - 2k $ to represent the end of the interval. 

For example, let $b_{i}^{(1)}, \dots, b_j^{(K)}$ be the optimal solution, (optimal time zone borders), i.e.\ they represent the optimal $K$-level quantizer for the interval $(i, M_K)$. With $j^{\prime} < j$, $b_{i}^{(1)}, \dots, b_{j^{\prime}}^{(K-1)}$ must represent the optimal $(K-1)$-level quantizer for the interval $(i,{M_{K-1}})$. Thus, we can split the problem into subproblems as:
\begin{equation}
\begin{split}
    \mathcal{D}_k(b_i, b_{M_k}) = \min_{\substack{\alpha_m}}  \mathcal{D}_1(b_i, b_m) + \mathcal{D}_{k+1}(b_m, b_{M_{k+1}}) \\ 
    \textrm{s.t.}\quad m \in (i, {M_k})
    \mbox{,}
    \label{eq:dk}
\end{split}
\end{equation}
such that the longitude line $b_m$ is between lines $b_i$ and $b_{M_k}$. The value $\alpha_m$ is the minimum distortion attained within the interval $(i, M_k)$ for time zone $\mathcal{B}_k$. Therefore, the optimal time zone border $ b_m^{(k)}$ for time zone region $\mathcal{B}_k$ is
\begin{equation}
    b_m^{(k)} = \argmin\limits_{b_m} \mathcal{D}_k(b_i, b_{M_k}) \mbox{.}
\end{equation}
We solve \eqref{eq:dk} up until $k+1 \leq K$. We follow a similar procedure to find optimal representation points $r_n^{(k)}$ by replacing $b_i$ with $r_{i-1}$ in \eqref{eq:d1} and \eqref{eq:dk}. 

The DP algorithm is shown in Algorithm \ref{algo:dp}. Given the inductive nature of the process, computations must start from $j=2$. In line 15 in Algorithm \ref{algo:dp}, we iteratively consider each longitude line $b_i$ as the prime meridian and apply DP, based on \eqref{eq:dk}. Then, the longitude line $b_m^{(k)}$ is the boundary of time zone $k$. 

\begin{algorithm}
\caption{Quantization with DP}\label{algo:dp}
\kwInput{$N$, $b_i$ for $i=0$ to $N$, $K$}
\KwResult{$b_{m^{\ast}_k}$ for all $k \in [0, K]$}
Compute and store $\mathcal{D}_1(b_i, b_j)$ for all $i, j \in [0, N]$ \Comment*[r]{by Eq. \eqref{eq:d1}}
$\Tilde{N} = N$ \\
\SetKwFunction{FMain}{DP}
\SetKwProg{Fn}{Function}{:}{}
\Fn{\FMain{$i$,$k$,$\Tilde{N}$}}{  
$v_{min} \gets \infty$ \\
$k \gets 0$ \\
compute $M_k = \Tilde{N}-2k$ \\
\For{$j\gets 2$ \KwTo $M_k$}{
$v \gets \mathcal{D}_1(b_i, b_j)$ \\
$v \gets v + $\FMain($i+1, k+1,\Tilde{N}-j$) \\ 
    \If{$v < v_{min}$}
    {
    $v_{min} \gets v$ \\ 
    $m^{k}\gets i+j$ \\ 
    }
}
\textbf{return} $v_{min}, m^k$ 
}
\For{$i\gets 0$ \KwTo $N$}
{
$k = 0, \Tilde{N}=N$ \\ 
$v, m_{k} \gets $\FMain($i, k,\Tilde{N}$)
}
Returns the set of longitude lines $\{b^{(1)}_m  \dots b^{(k)}_{m^{\prime}} \dots b^{(K)}_{m^{\prime\prime}} | \quad \forall k \in [1,K]\}$, which are the optimal time zone boundaries. 
\end{algorithm}
\subsection{Reinforcement Learning-Based DP}
The computational complexity of the DP approach makes computation challenging for large values of $K$ and $N$, e.g.\ $K=24$. In DP, to solve \eqref{eq:dk}, we have a total of $K\times N^2$ tuples for a given two points, each tuple needs to be computed at most $N$ times. For the total set of points, \eqref{eq:dk} is solved $N$ times. Therefore we end up with a worst-case computation of $\mathcal{O}(N(KN^3))$. 

\subsubsection{Markov Decision Process}
Given the dependency in finding partitions in \eqref{eq:dk}, the variables at $k$-level quantizer depend on future $(k+1)$-level quantizer results. Therefore, the quantization problem for partitioning longitude lines can be cast as a Markov Decision Process (MDP). 

MDP is a controlled Markov chain, in which the transition from one state to another depends on an external control parameter called the \textit{action}. Specifically, the probability of transitioning to state $s^{\prime}$ from state $s$ upon taking action $a$ is denoted by $P_{s,s^{\prime}}(a)$ and given by 
\begin{equation}
\begin{split}
    \mathcal{P}(s,s^{\prime},a) := P(s_1 = s^{\prime} | s_0 = s, A_0 = a)\mbox{, } \\ \forall s, s^{\prime} \in \mathcal{S}, \forall a \in \mathcal{A}(s)\mbox{ .}
    \label{eq:action_prop}
\end{split}
\end{equation}
In addition to the finite state space $\mathcal{S}$, for any state $s\in \mathcal{S}$, MDP has a finite action space $\mathcal{A}(s)$ of possible actions that can be taken at state $s$. Another component of an MDP is an underlying one-step \textit{reward} function $\mathcal{R}(s,a)$ that assigns random rewards to each $(s,a)$ pair. Therefore, an MDP is given by a quadruplet $(\mathcal{S}, \mathcal{A}, \mathcal{P}, \mathcal{R})$. The goal in analyzing an MDP is to find the optimal \textit{policy} $\pi^{\ast}_t$ at different time steps. More formally, a policy is a probability distribution over the action space $\mathcal{A}$:
\begin{equation}
        \pi_t(a | \{s_i, a_i\}_{i=0}^{t-1}, s_t)  :=  P(\mathcal{A}(s_t) = a | s_0, a_0, \dots, s_{t-1}, a_{t-1}, s_t) \mbox{.}
\end{equation}
That is, the current policy $\pi_t$ at current time $t$ depends on a sequence of previous states and actions from $t=0$ up to $t=t-1$ and current state $s_t$. 

In solving large MDPs, there are three challenges \cite{powell2007approximate}: (i) the size of the state space $\mathcal{S}$ may be too large to evaluate within a reasonable time, (ii) the size of actions may be too large to find the optimal action for all states within a reasonable time, and (iii) computing the expectation of \textit{future} costs may be intractable when levels space is large, e.g., $k$-level depends on future $(k+1)$-level quantizer. Given such challenges, approximate dynamic programming (ADP) is an alternative. The output of ADP is a policy or decision function that maps each possible state $s_t$ to a decision $\pi_t$ at each stage $t$ in the time horizon. ADP combines optimization with simulation to approximate the optimal solution of MDPs \cite{mes2017approximate}. ADP is a form of reinforcement learning (RL).  

\subsubsection{Value Iteration}
One RL technique to solve an MDP is the \textit{value iteration} (VI) algorithm, where all states are updated in random order at one iteration. It is also based on a one-step look-ahead search from the current state \cite{sutton2018reinforcement}. 

The VI function $\mathcal{V}(\cdot)$ basically uses the Bellman optimality equation \cite{bellman1966dynamic} with an update rule at each iteration from $t=1$ to $T$. Each state $s \in S$ has a value $\mathcal{V}_t(s)$ that is updated using the previous value function of the next state $s^{\prime} \in S$ with a non-negative reward function $R(s,a)$ to dictate the best action to take over all possible actions $a \in \mathcal{A}(s)$ as in  \eqref{eq:action_prop}. The reward indicates what is the good action to take in an immediate sense, whereas a value function indicates what is the best in the long run. Roughly speaking, the value of a state is the total amount of reward accumulated over the future, given the current state \cite{sutton2018reinforcement}. Formally, the value function iterates from $t$ to a total of $T$ and is expressed as follows,  
\begin{equation}
    \mathcal{V}_t(s) = \max\limits_{a \in \mathcal{A}(s)} \mathcal{R}(s,a) + \gamma  \sum\limits_{s^{\prime} \in \mathcal{S}}  \mathcal{P}(s,s^{\prime},a) \mathcal{V}_{t-1}(s^{\prime}) \mbox{.}
    \label{eq:vi_original}
\end{equation}
The parameter $\gamma$ is a \textit{discount} factor, $0 < \gamma < 1$ which is used to make an infinite sum finite when $T \to \infty$. Therefore, this also imposes convergence guarantees as proved in \cite{sutton2018reinforcement,agarwal2019reinforcement}. Specifically, running the value function in  \eqref{eq:vi_original} for $T$ iterations such that 
\begin{equation}
     T \geq \frac{\log(||\mathcal{V}_1 - \mathcal{V}_0||_{\infty})+\log(2)  - \log(\epsilon(1-\gamma))}{\log(\frac{1}{\gamma})}
     \label{eq:Tlower_bound}
\end{equation}
ensures that the optimal value function $\mathcal{V}^{\ast}$ is $||\mathcal{V}_{T} - \mathcal{V}^{\ast}||_{\infty} < \epsilon$, therefore, $\pi_T \to \pi^{\ast}$, the optimal policy. For the theorems and derivations that lead to this result, please see \cite{agarwal2019reinforcement, Csaba2020, Srikant2022}

\subsubsection{Quantization with Value Iteration}
To convert our  DP function \eqref{eq:dk} into a VI function, we consider redefining the states, rewards, and actions based on our setting. First, we rewrite \eqref{eq:vi_original} with equal transition probabilities and with a minimization objective as follows:
\begin{equation}
    \mathcal{V}_t(s) = \min\limits_{a \in \mathcal{A}(s)} \mathcal{R}(s,a) + \gamma  \mathcal{V}_{t-1}(s^{\prime}) \mbox{.}
    \label{eq:vi_2}
\end{equation}
Then, we build the MDP using the set of longitude lines $\mathcal{B}$ and the number of time zones $K$ as follows 
\begin{equation}
    \begin{split}
        s & := (b_i, k, b_{M_k}) \\
        a & := b_m \\
        \mathcal{A}(s) & := f(b_i, k, b_{M_k}) \\
        \mathcal{R}(s,a) & := \mathcal{D}_1(b_i, b_m) \\ 
        s^{\prime} & := (b_m, k+1, b_{M_{k+1}} )\mbox{.} 
    \end{split}
\end{equation}
The function $f(b_i, k, b_{M_k})$ returns actions $\mathcal{A}(s)$ that satisfy the constraints on $b_m$ in \eqref{eq:dk}. 

Therefore, combining  \eqref{eq:dk} and  \eqref{eq:vi_2} with change in notation we get
\begin{equation}
    \mathcal{V}_t(b_i, k, b_{M_k}) = 
    \min\limits_{b_m \in \mathcal{A}(b_i, k, b_{M_k})} \mathcal{R}(b_i,b_m) + \gamma \mathcal{V}_{t-1}(b_m, k+1, b_{M_{k+1}} ) \mbox{.}
\end{equation}
Notice that if $V_t$ converges before $T$, that is $\mathcal{V}_{T-1} = \mathcal{V}_{T}$, then it satisfies the DP in  \eqref{eq:dk}
\begin{equation}
    \mathcal{V}_T(b_i, k, b_{M_k}) = 
    \min\limits_{b_m \in \mathcal{A}(b_i, k, b_{M_k})} \mathcal{R}(b_i,b_m) + \gamma \mathcal{V}_T(b_m, k+1, b_{M_{k+1}} ) \mbox{.}
\end{equation}

The complete algorithm is given in Algorithm \ref{algo:rldp}. We  run the function following the lower bound in \eqref{eq:Tlower_bound}. After convergence, the optimal policy $\pi^{\ast}$ is the set of longitude lines $\{b_m^{(1)}, \dots, b_{m^{\prime}}^{(K)}\ | \ m, m^{\prime} \in [0,N]\}$ which are the optimal boundaries for each time zone $\mathcal{B}_k$, (each boundary with a different index $m \in [0,N]$), and their corresponding representation points $\{r_n^{(0)}, \dots, r_{n^{\prime}}^{(K-1)}\ | \ n, n^{\prime} \in [0,N]\}$ (each at different index $n \in [0,N]$). 

\begin{algorithm}
\caption{Quantization with VI}\label{algo:rldp}
\kwInput{$N$, $b_i$ for $i=0$ to $N$, $K$}
\KwResult{$\{m^{\ast}_{k}, \ k \in [0, K], \ m \in [0, N]\}$}
Compute and store $D_1(b_i, b_m)$ to be $\mathcal{R}(b_i,b_m)$ for all $i, m \in [0, N]$ \Comment*[r]{by Eq. \eqref{eq:d1}}
Build MDP $(\mathcal{S}, \mathcal{A}, \mathcal{R})$ from $N$, $b_i$ and $K$ \\ 
Initialize $\mathcal{V}_0$ for all $s \in \mathcal{S}$ \\
Initialize $T$ \\
\For{$t\gets 0$ \KwTo $T$}
{
    \For{$s = (b_i, k, b_{M_k}) \in \mathcal{S}$}
    {
        $\mathcal{V}_t(b_i, k, b_{M_k}) = \min\limits_{b_m \in \mathcal{A}(b_i, k, b_{M_k})} \mathcal{R}(b_i,b_m) + \gamma \mathcal{V}_{t-1}(b_m, k+1, b_{M_k} )$ \\
    }
}
$\pi^{\ast} \gets \{m^{k}\} \forall k \dots K$. The longitude lines at indices $m$ for all $k \in K$, gives the optimal time zone boundaries.  
\end{algorithm}

\section{Empirical Demonstration}
\label{sec:empirical}
In this section, we demonstrate our approach for the time zone design problem so as to explore the possibility of improving social capital, social mobility, and COVID-19 health.  The econometric findings on the causal relationship between sunset time and these measures of human wellbeing that emerge in the RDD analysis may be of independent interest.

\subsection{Data and Pre-processing for Regression Discontinuity Models}
Let us initially consider the continental United States.  Following \cite{Giuntella_2019}, we consider the distance between the centroid of a region and the time zone boundary to calculate the daylight hours for a given region. We use Census center of population and time zone boundary data from the Bureau of Transportation Statistics to compute the distance between the centroid of the county and adjacent time zone boundaries. 
\paragraph{Census Centers of Population} The Census centers of population dataset\footnote{\url{https://www.census.gov/geographies/reference-files/time-series/geo/centers-population.html}} provides the balance point of various geographic and demographic features. We especially use the centers of the population by county data to obtain coordinates of counties in the continental U.S. With this dataset, we compute the average sunset time given year and distance to the time zone boundary for all counties. 
\paragraph{Time Zone Boundary} U.S.\ time zone boundaries and daylight saving time (DST) are managed by the Department of Transportation. There are four time zones in the continental U.S.: Pacific (UTC $-$07:00), Mountain (UTC $-$06:00), Central (UTC $-$05:00) and Eastern (UTC $-$04:00). There is a one hour difference between each time zone. As time zones in the U.S.\ are not strictly based on mean solar time at the meridian, we use the shape file provided by the Bureau of Transportation Statistics.  With coordinates of counties from Census data, we compute the distance between adjacent time zone boundaries and centroids of counties using Euclidean distance. Since not all states observe DST or have a consistent time zone over years, we exclude counties in Arizona, Florida, and Indiana from our analysis. The distance from the time zone boundary and the current year are used to calculate the average sunset time for a given region. 
\paragraph{Social Capital} The social capital measure is obtained from the Social Capital Project \cite{sc_project} and comprises \textit{family unity}, an indicator of the structure of families in terms of marriage and children; \textit{community health}, an indicator of participation in civic life; \textit{institutional health} that considers confidence in media/corporations/schools and participation in institutions such as elections and census; and \textit{collective efficacy}, an indicator for the converse of social disorganization, operationalized via violent crime rates.
\paragraph{Social Mobility} The social mobility data is from \cite{Chetty_2014}, which investigates intergenerational mobility in the United States. We focus on absolute upward mobility, which indicates the expected rank of children whose parents are at the 25th percentile of the national income distribution based on rank-rank regression, and relative upward mobility which is the slope from linear regression of child rank on parent rank.

\paragraph{COVID-19 Cases and Fatalities} We use the cumulative counts of COVID-19 cases and deaths in the United States from the New York Times \cite{nytiems_cov}. It contains data from the first reported COVID-19 case in Washington state on January 21, 2020. Since the data shows much heterogeneity, we take the log of the cases and fatalities (adding 1 to all data points to avoid the value 0 in the log).

\subsection{RDD Analysis Results}
Let us present the counterfactual prediction results of the effect of time zone borders on social capital, social mobility, and COVID-19 cases. 

\subsubsection{Sunlight discontinuity} 
\begin{figure}[!ht]
  \centering
    \includegraphics[width=0.6\linewidth]{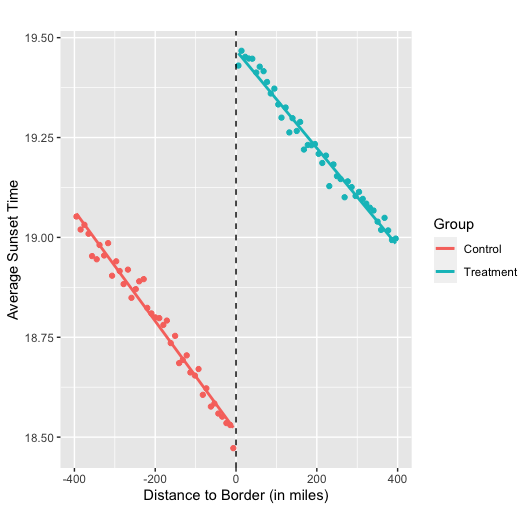}
    \caption{Average Sunset Time vs. Distance to Border (in miles) in Social Capital.}\label{fig:SC_Sunset}
\end{figure}

We show the difference of average sunset time to the east and west of a time zone border.  In Figure \ref{fig:SC_Sunset}, we see a sharp discontinuity on the time zone boundary. At the boundary (i.e., the distance to border is 0), there is approximately a one hour difference in average sunset time between eastern and western counties. Counties located to the eastern side of time zone boundary have later average sunset time, which results in less amount of sleep, whereas counties located to the western side of the time zone boundary have earlier average sunset time.  This effect increases with distance from the time zone boundary.

\subsubsection{Counterfactual for Social Capital}
\label{sec:sc}
Social capital includes five different metrics: family unity, community health, institutional health, efficacy, and an overall index that averages the four sub-indices. Here we investigate the effect of sunlight amount determined by time zone borders on social capital across counties. 

Table \ref{Tab:Social_Capital} reports the local non-parametric RDD estimates for outcomes of interest. This table shows the results of conventional and robust local non-parametric regression discontinuity estimates with uniform kernel developed by \cite{Calonico_2014}. We use MSE-optimal bandwidth and manipulation testing using local polynomial density estimation proposed by \cite{Calonico_2014} and \cite{Cattaneo_2018}.  

We find a significant causal effect from sunset time on both the overall social capital index and the community health sub-index for both conventional and robust models. We see the areas where people sleep less tend to have less social capital and less community health. For institutional health, the treatment effect is not significant in the conventional model, but it is significant in the robust model at a $0.1$ significance level. 

On the other hand, the treatment effect for efficacy is significant in the conventional model, and not significant in the robust model. It indicates we may not see any causality between sleep and institutional health in the local fit, meanwhile, there is a (weak) causal link between sleep and institutional health in the global fit. Similarly, the causality between sleep and efficacy is only meaningful given bandwidth. Lastly, family unity does not have any significant causal link with sleep.

\begin{table}
\caption{Social Capital: Local non-parametric regression discontinuity estimates.}
\label{Tab:Social_Capital}

\scalebox{0.75}{
\begin{tabular}{l|ll|l|l}
\hline
 & \multicolumn{2}{c|}{Later Sunset Counties} & \begin{tabular}[c]{@{}l@{}}Bandwidth \\ (in miles)\end{tabular} & Manipulation \\ \hline
 & \multicolumn{1}{l|}{Conventional} & Robust &  &  \\ \hline
County Index & \multicolumn{1}{l|}{$-0.522^{***} (0.199)$} & $-1.194^{***}(0.419)$ & 208.792 & No \\ \hline
Family Unity & \multicolumn{1}{l|}{$-0.204 (0.251)$} & $-0.491 (0.540)$ & 194.451 & No \\ \hline
Community Health & \multicolumn{1}{l|}{$-0.790^{***} (0.274)$} & $-1.644^{***}(0.558)$ & 200.494 & No \\ \hline
Institutional Health & \multicolumn{1}{l|}{$-0.080 (0.198)$} & $-0.694^{*}(0.395)$ & 223.144 & No \\ \hline
Efficacy & \multicolumn{1}{l|}{$-0.706^{***}(0.225)$} & $-0.519(0.447)$ & 206.011 & No \\ \hline
\end{tabular} 
}

\begin{minipage}{\columnwidth}
Note: In the later sunset counties row, first number represents the estimate, and \* shows the significance levels where $^{*}p<0.1$, $^{**}p<0.05$, and $^{***}p<0.01$, and number in parenthesis displays the standard error.
\end{minipage}
\end{table}

\begin{figure}[t]
    \centering
    \includegraphics[width=0.65\linewidth]{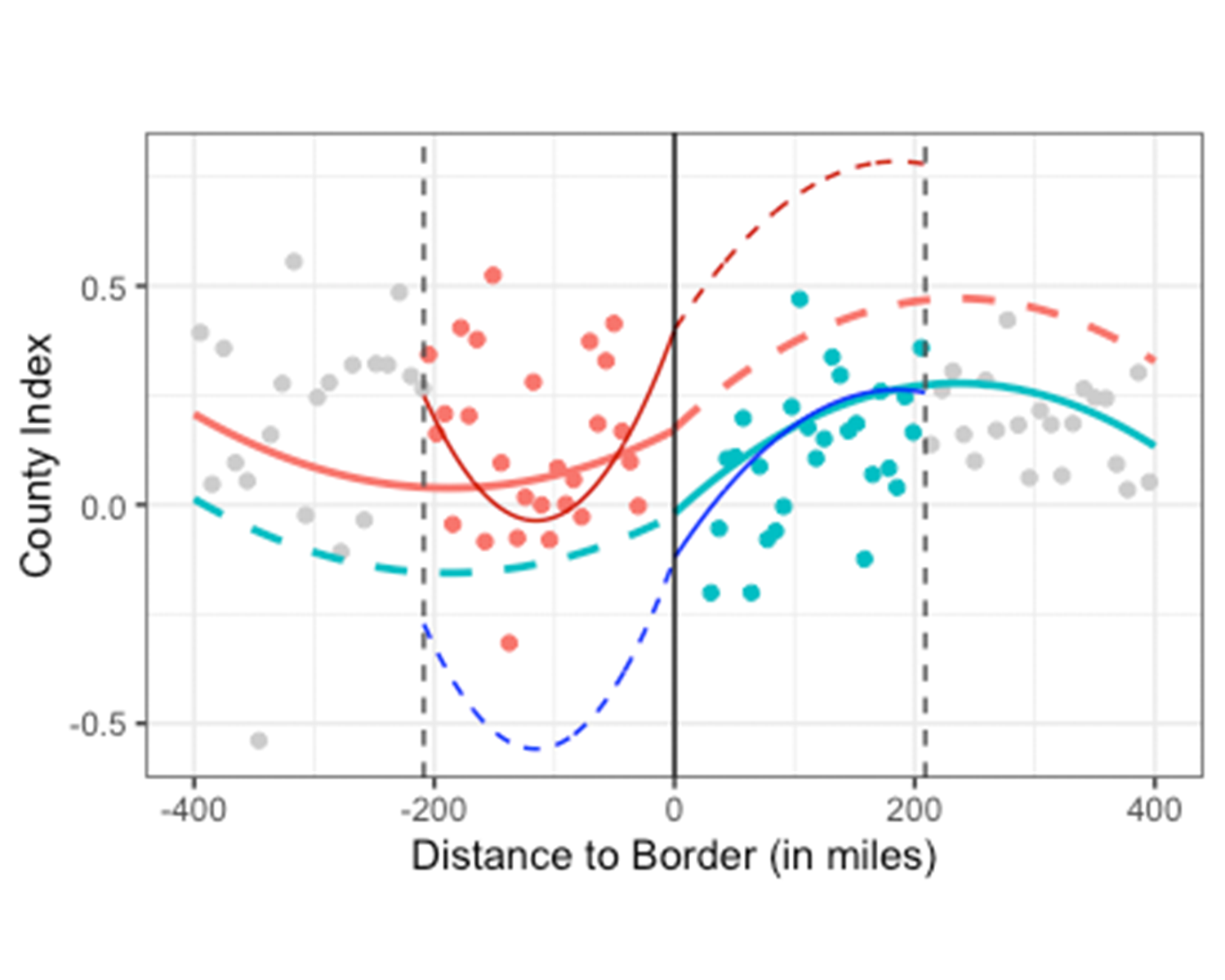}
    \caption{The causal estimation of social capital (Pink line for the global fit of control group, light blue line for global fit of treatment group, red line for local fit of control group, blue line for local fit of treatment group, each dashed colored line represents the estimation with continuity assumption, while dashed gray vertical bars shows the bandwidth used for local fit, and black solid line indicates the cut-off).}
    \label{fig:social capital}
\end{figure}

Figure \ref{fig:social capital} shows counterfactual lines for the overall social capital index outcome. The social capital index has both conventional and robust model significance; the counterfactual lines of local and global fit behave similarly. We find that later sunset causes a lower social capital index. Thus, if the western counties near the time zone boundary were moved to the eastern side of the time zone boundary, we should expect a lower counterfactual prediction of the county index. However, both local (red dashed line) and global (pink dashed line) imply their county indices would be higher near the eastern side of the boundary. 

\subsubsection{Counterfactual for Social Mobility}
We focus on absolute and relative mobility. Like the previous social capital analysis, we observe the same pattern of average sunset time of counties at each side of the time zone boundary on social mobility. This is because the average sunset time does not change much over different years. Table \ref{tab:Social_Mobility} shows the local non-parametric RDD estimates for absolute and relative mobility.
We see there may not be strong causality between sleep and social mobility. In absolute mobility, the treatment effect from the conventional model is only significant at a $0.1$ significance level, and the estimate from the robust model is not significant. Neither conventional nor robust models in relative mobility are statistically significant.

\begin{table}
\caption{Social mobility: Local non-parametric regression discontinuity estimates
}

\scalebox{0.80}{
\begin{tabular}{l|ll|l|l}
\hline
 & \multicolumn{2}{c|}{Later Sunset Counties} & \begin{tabular}[c]{@{}l@{}}Bandwidth\\ (in miles)\end{tabular} & Manipulation \\ \hline
 & \multicolumn{1}{l|}{Conventional} & Robust &  &  \\ \hline
Absolute Mobility & \multicolumn{1}{l|}{$-2.652^{*} (1.471)$} & $-1.653(3.000)$ & 235.399 & No \\ \hline
Relative Mobility & \multicolumn{1}{l|}{$-0.012(0.009)$} & $-0.002 (0.019)$ & 220.851 & No \\ \hline
\end{tabular}
}

\begin{minipage}{\columnwidth}
Note: In the later sunset counties row, the first number represents the estimate, and \* shows the significance levels where $^{*}p<0.1$, $^{**}p<0.05$, and $^{***}p<0.01$, and number in parenthesis displays the standard error.
\end{minipage}
\label{tab:Social_Mobility}
\end{table}

\begin{figure}[!ht]
  \centering
\includegraphics[width=0.85\linewidth]{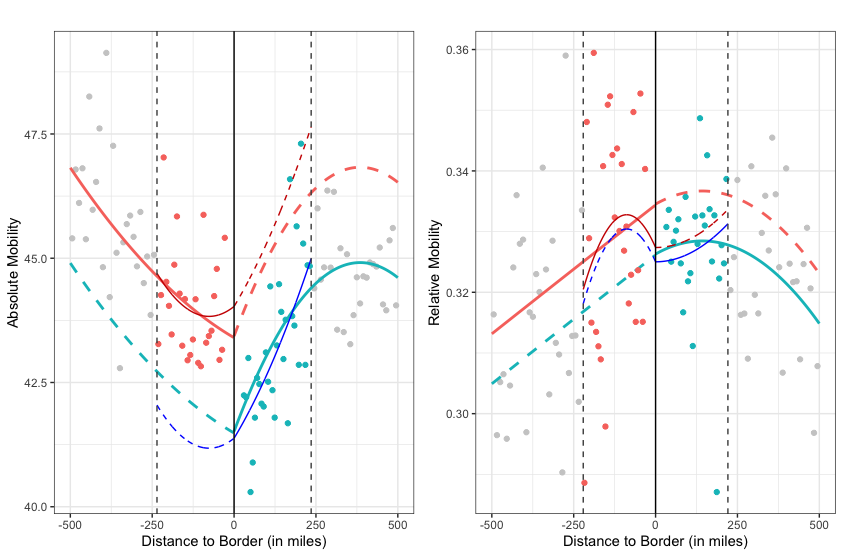}
\caption{Counterfactual Prediction in Social Mobility. (The pink line shows the global fit for the control group, the light blue line for the global fit for the treatment group, the red line for the local fit for the control group, blue line for the local fit for the treatment group. Each dashed line represents the counterfactual prediction with continuity assumption. Dashed gray vertical bars mean the bandwidth to be used for local fit, and the black solid line indicates the cut-off.)}
\label{fig:SM_predict}
\end{figure}

Figure \ref{fig:SM_predict} displays the counterfactual lines in social mobility. As there was no causal relationship between sunset and social mobility, the counterfactual lines may not be meaningful here. Although the discontinuity appears visually at the time zone boundary, it is not statistically significant (i.e., social mobility is not notably different between control and treatment groups). 

\subsubsection{Counterfactual for COVID-19 Cases and Fatalities}
We investigate the causality between sunset and the log of COVID-19 cases and fatalities. In Table \ref{Tab:COVID}, we see there is a strong causal link between sunset time and COVID-19 cases and fatalities. We observe that the counties at the eastern side of time zone borders tend to have fewer cases and fatalities.  This may be largely due to social factors \cite{FerwanaV2021}, but there might also be a biological contribution, e.g.\ mediated through melatonin \cite{Zhou_2020}.

\begin{table}

\caption{COVID-19 Cases and Fatalities: Local non-parametric regression discontinuity estimates.}

\scalebox{0.80}{
\begin{tabular}{l|ll|l|l}
\hline
 & \multicolumn{2}{c|}{Later Sunset Counties} & \begin{tabular}[c]{@{}l@{}}Bandwidth \\ (in miles)\end{tabular} & Manipulation \\ \hline
 & \multicolumn{1}{l|}{Conventional} & Robust &  &  \\ \hline
$\log(\text{case})$ & \multicolumn{1}{l|}{$-0.413^{***}(0.092)$} & $-0.536^{***} (0.174)$ & 297.753 & No \\ \hline
$\log(\text{fatalities})$ & \multicolumn{1}{l|}{$-0.425^{***}(0.112)$} & $-0.437^{**}(0.203)$ & 270.199 & No \\ \hline
\end{tabular}
}

\begin{minipage}{\columnwidth}
Note: In the later sunset counties row, the first number represents the estimate, and \* shows the significance levels where $^{*}p<0.1$, $^{**}p<0.05$, and $^{***}p<0.01$, and number in parenthesis displays the standard error.
\end{minipage}

\label{Tab:COVID}
\end{table}

\begin{figure}[!ht]
  \centering
  \includegraphics[width=0.85\linewidth]{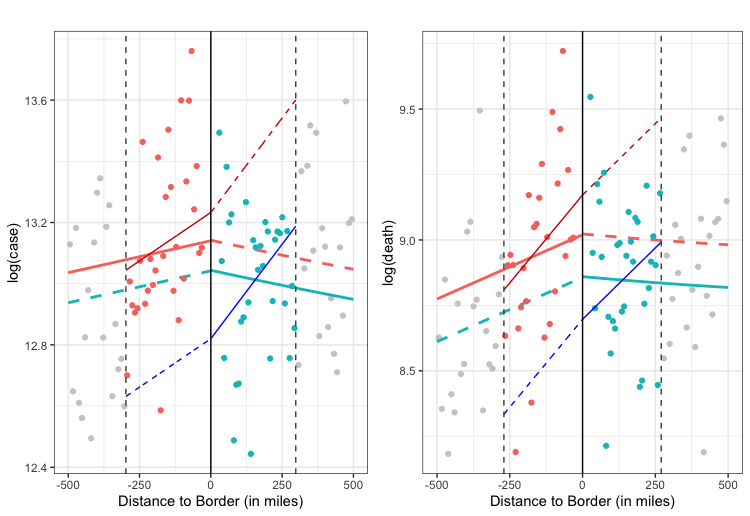}
  \caption{Counterfactual Prediction in Social Mobility. This figure shows the counterfactual prediction of the outcome of interests in social capital. (Pink line shows the global fit for the control group, the light blue line for the global fit for the treatment group, the red line for the local fit for the control group, blue line for the local fit for the treatment group. Each dashed line represents the counterfactual prediction with continuity assumption. Dashed gray vertical bars mean the bandwidth to be used for local fit, and the black solid line indicates the cut-off).}\label{fig:COVID_predict}
\end{figure}

In Figure \ref{fig:COVID_predict}, we display the counterfactual lines. We see there is a strong disagreement between global and local fit in both COVID-19 cases and fatalities. In particular, local counterfactual lines of COVID-19 case go in the opposite direction as the result of causal inference. Also, global counterfactual line of treatment group goes downward, contrary  to our intuition. This pattern also appears for COVID-19 fatalities. This is strong evidence that we cannot make a counterfactual prediction only with the continuity assumption.

\subsection{Quantization}
Given our finding of causal significance between sunset and social capital in Section~\ref{sec:sc}, we focus on those indices to demonstrate our quantization approach. 

To partition regions, we base our partitioning on the population size at each longitude line \cite{Rankin2008}. We use the world population estimates at the west and east of each of the 360 longitude lines, with 15 degrees from both sides of a longitude line, hence, population sizes are observed at a total of 720 longitude coordinates/points. We use population data \cite{popData2000} calculated at each longitude line of the year 2000. At each time zone region, we aim to have a minimum population size at the east edge of a time zone border to minimize the distortion effect following \eqref{eq:prime_meridian}--\eqref{eq:k_opt}.

\subsubsection{Uniform Quantization}
\begin{figure}[!ht]
    \centering
    \includegraphics[width=0.90\linewidth]{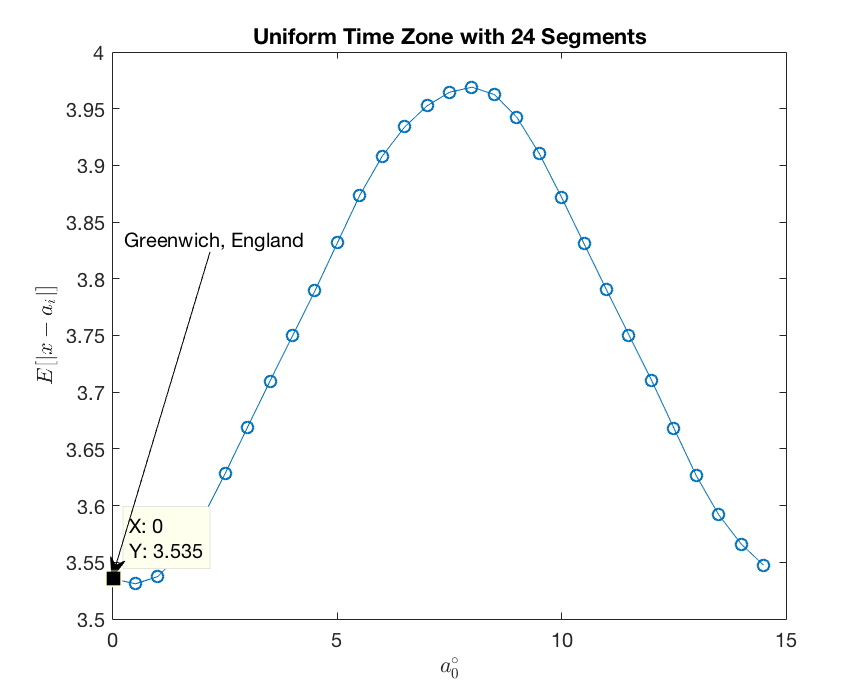}
    \caption{The distortion in circadian rhythm effect given uniform quantization with $K=24$.}
    \label{fig:uniformTZ1}
\end{figure}

\begin{figure}[!ht]
    \centering
    \includegraphics[width=0.90\linewidth]{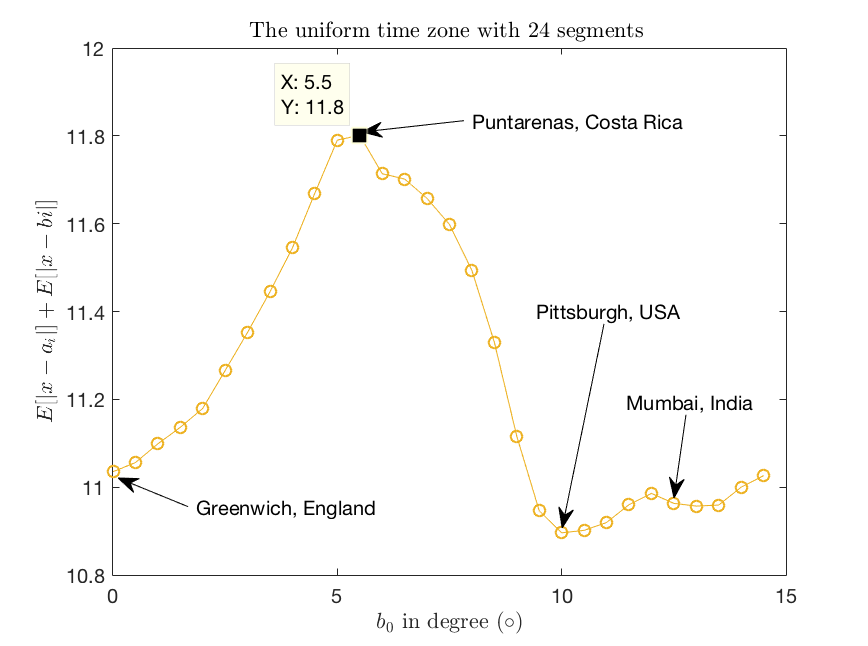}
    \caption{The total distortion in sun effect at 30 quantized regions  with $K=24$.}
    \label{fig:uniformTZ2}
\end{figure}

We investigate the east edge effect and sun effect at each partition. For a given Prime Meridian in Greenwich, the distortion in east edge effect is calculated under 24 time zones for uniform partitions. Figure \ref{fig:uniformTZ1} shows the distortion in circadian rhythm at each time zone segment with uniform zone width, as distortion increases for regions away from the Prime Meridian. Figure \ref{fig:uniformTZ2} shows the distortion in east edge effect in addition to the sunlight amount effect. 

\begin{figure}[!ht]
    \centering
    \includegraphics[width=1\linewidth]{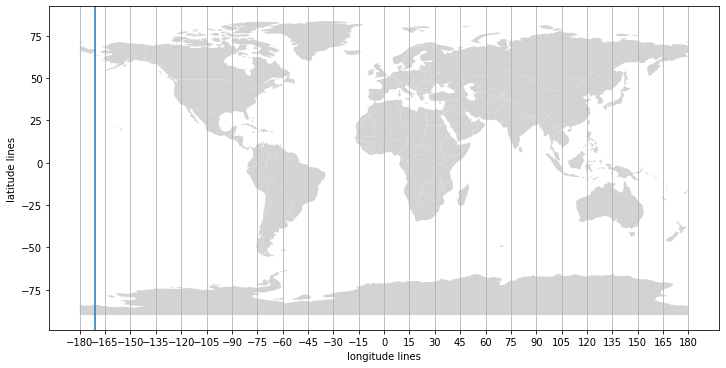}
    \caption{The optimal Prime Meridian for fixed-sized time zones under $K=24$. Optimal prime meridian in blue}
    \label{fig:uniform_prime}
\end{figure}

For the Prime Meridian choice problem, we apply our method of quantization with keeping the size of time zone regions fixed. Figure \ref{fig:uniform_prime} shows the selected Prime Meridian at the longitude line 170.75 West. 

\subsubsection{Quantized Time Zone Borders}
\begin{figure}[!ht]
    \centering
\includegraphics[width=0.98\linewidth]{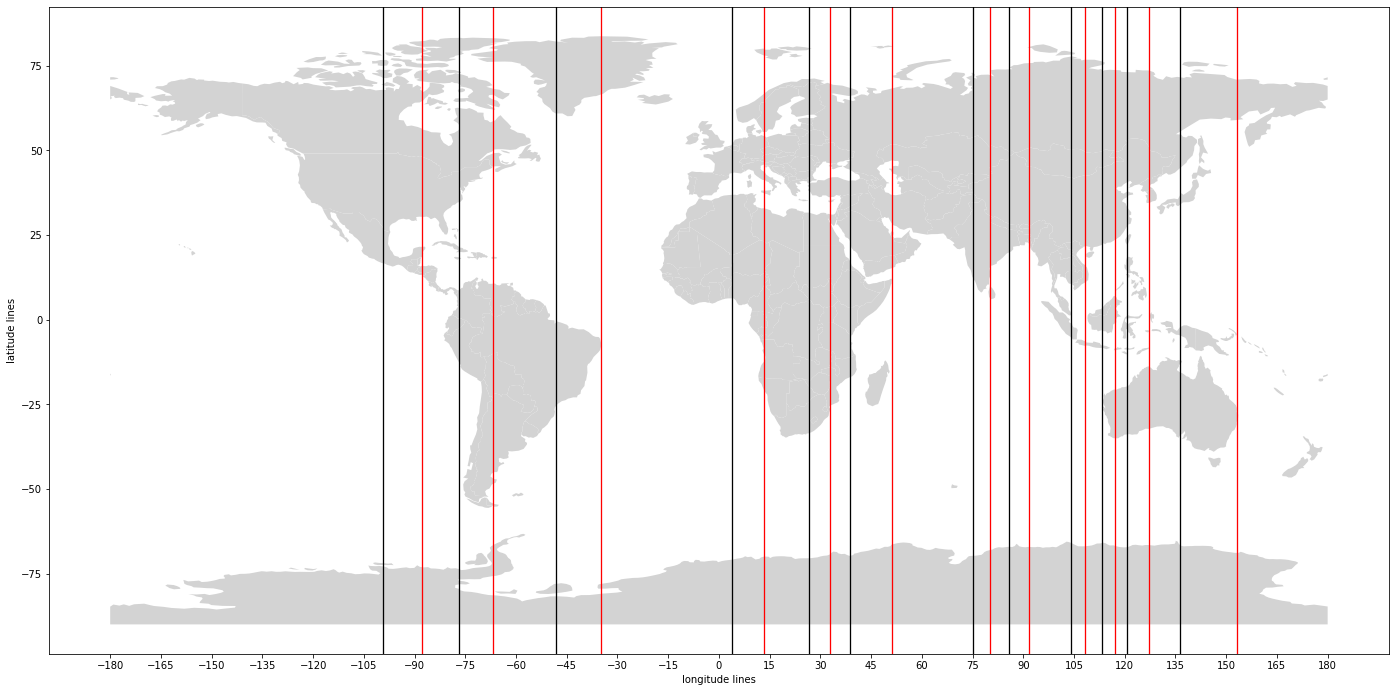}

    \caption{Time zone quantization for $K=12$. (Black lines are the boundaries of a time zone and the red lines are the representation points) }
    \label{fig:tz12}
\end{figure}

\begin{figure}[t]
    \centering
    \subfloat[Quantized World partitions\label{fig:tz24}]{
\includegraphics[width=1\linewidth]{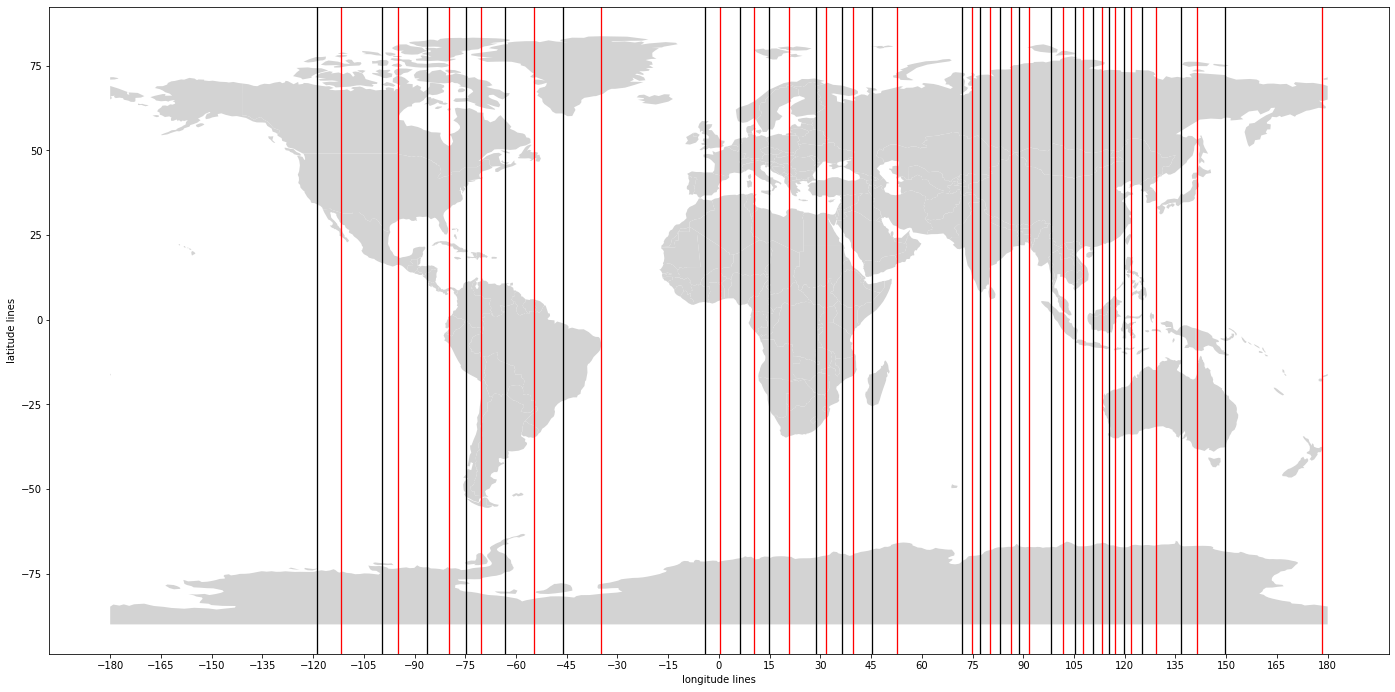}}
\vfill
    \subfloat[Quantized Continental United States partitions\label{fig:USTZ}]{
    \includegraphics[width=1\linewidth]{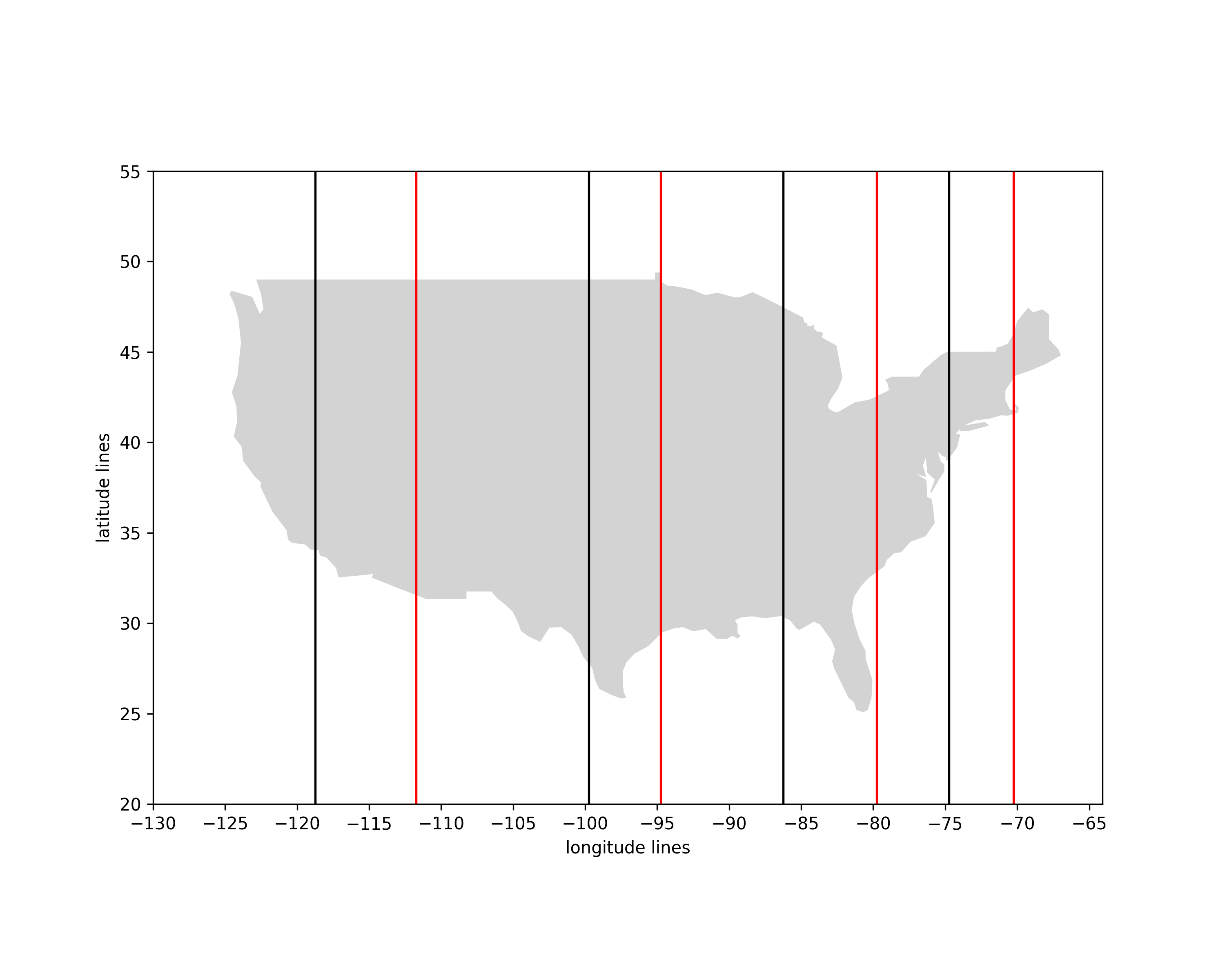}}
    \caption{Time zone quantization for $K=24$. (Black lines are the boundaries of a time zone and the red lines are the representation points)}
    \label{fig:all_tz24}
\end{figure}

We apply our method in Algorithm \ref{algo:rldp} and investigate the quantized partitioning by allowing the time zone regions to be of non-uniform widths. Figures \ref{fig:tz12} and \ref{fig:tz24} show the optimal time zone boundaries redesigned by quantization over the world map. We consider the number of time zones to be $K=12$ and $K=24$ respectively. We see that more time zone boundaries appear just to the east of regions with large populations, e.g.\ India and China. This follows our intuition for the need to redesign optimal time zone boundaries that minimize distortions across populations. 

\subsection{Designed Discontinuity Counterfactual Prediction}

After we have optimally redesigned time zone borders as in Figure~\ref{fig:tz24}, with a focus on the United States in Figure~\ref{fig:USTZ}, we measure the counterfactual change in the causal impact of the new borders on social capital, as compared to the impact under the standard time zone borders. Focusing on the continental United States, we use the same social capital data \cite{sc_project} and reuse the RDD model in \eqref{eq:rd}, following the same causal inference analysis in Section \ref{sec:RD_setting}, with the robust RDD model design. Counties are assigned to a time zone region with the minimum distance from a time zone border. Counties that are to the east of the time zone border have $h_c = 1$, to indicate the treatment, and hence, measure the treatment effect $\beta_1$ in \eqref{eq:rd}. The outcome of interest, $Y_c$, is the social capital index for county $c$, including the aggregated measure, county index, and subindex measures. 

Table ~\ref{tab:counterfactual_quantization} shows the effect of being at the eastern edge of the redesigned time zone border on social capital indices. In comparison to results in Table \ref{Tab:Social_Capital}, we see that the quantized borders have yielded a stronger positive causal impact on social capital which mitigates the effect of distortion from the standard time zone borders. For example, the effect size changed from $-1.194$ to $0.083$ for the county index indicating a positive effect of the redesigned time zone borders. Similarly, the community health index shows a smaller effect in comparison to the standard time zone borders effect.  

\begin{table}
\caption{Designed Discontinuity Counterfactual Prediction on Social Capital: Local non-parametric regression discontinuity estimates}

\centering
\begin{tabular}{l|l|l}
\hline
 & Later Sunset Counties (Robust) & Manipulation \\ \hline
County Index & $0.083^{***}(0.024)$ & No \\ \hline
Family Unity & $0.098^{***}(0.022)$ & No \\ \hline
Community Health & $-0.0443^{**}(0.019)$ & No \\ \hline
Institutional Health & $0.073^{**}(0.024)$ & No \\ \hline
Efficacy & $0.047^{**}(0.019)$ & No \\ \hline
\end{tabular}

\begin{minipage}{\columnwidth}
Note: In the later sunset counties row, first number represents the estimate, and \* shows the significance levels where $^{*}p<0.1$, $^{**}p<0.05$, and $^{***}p<0.01$, and number in parenthesis displays the standard error.
\end{minipage}

\label{tab:counterfactual_quantization}
\end{table}
\section{Discussion}
\label{sec:discussion}
If we want as many people as possible living at the eastern edges of their time zones so the sun sets earlier, they sleep more, and therefore have more social capital \cite{Putnam2000} (while ensuring the sun and the clock do not differ too much), can we design time zone boundaries to do so? For this problem and structurally-related ones that involve the design of discontinuities such as in health and education policies, we studied a problem formulation at the intersection of causal inference and quantization theory for the purpose of mechanism design.  This led to new mathematical developments in linking regression discontinuity counterfactuals with optimal quantization theory.  For the time zone problem specifically, results put time zone boundaries just to the east of large population centers, and we showed the possibility of significant gains in social capital.

There are natural questions of equity that arise through the design of discontinuities of the type we developed here.  Characterization is of interest for future research.

More broadly, there are several novel signal processing questions of interest that arise in the context of causal inference and causal design.

\section*{Acknowledgement}
We thank Anay Pattanaik and Razan Baltaji for valuable discussions and useful suggestions. 
\bibliographystyle{IEEEtran}
\bibliography{refs}
\end{document}